\documentclass[pdflatex, sn-nature-num]{sn-jnl}

\usepackage{graphicx}%
\usepackage{multirow}%
\usepackage{amsmath,amssymb,amsfonts}%
\usepackage{amsthm}%
\usepackage{mathrsfs}%
\usepackage[title]{appendix}%
\usepackage{xcolor}%
\usepackage{textcomp}%
\usepackage{manyfoot}%
\usepackage{booktabs}%
\usepackage{algorithm}%
\usepackage{algorithmicx}%
\usepackage{algpseudocode}%
\usepackage{listings}%

\theoremstyle{thmstyleone}%
\theoremstyle{thmstyletwo}%

\theoremstyle{thmstylethree}%

\unnumbered%

\usepackage{comment}

\usepackage{siunitx}

\usepackage{cite}
\makeatletter
\renewcommand\@cite[2]{#1} 
\makeatother

\begin{document} 

\title[Experimental Verification of Electron-Photon Entanglement]{Experimental Verification of Electron-Photon Entanglement}

\author[1,2]{\fnm{Alexander} \sur{Preimesberger}}\email{alexander.preimesberger@tuwien.ac.at}
\equalcont{These authors contributed equally to this work.}

\author[1,2]{\fnm{Sergei} \sur{Bogdanov}}\email{sergei.bogdanov@tuwien.ac.at}
\equalcont{These authors contributed equally to this work.}

\author[1,2]{\fnm{Isobel C.} \sur{Bicket}}\email{isobel.bicket@tuwien.ac.at}

\author[1]{\fnm{Phila} \sur{Rembold}}\email{phila.rembold@tuwien.ac.at}

\author*[1,2]{\fnm{Philipp} \sur{Haslinger}}\email{philipp.haslinger@tuwien.ac.at}

\affil[1]{\orgdiv{VCQ, Atominstitut}, \orgname{Technische Universit\"{a}t Wien}, \orgaddress{\street{Stadionallee 2}, \city{1020 Vienna}, \country{Austria}}}

\affil[2]{\orgdiv{University Service Centre for Transmission Electron Microscopy}, \orgname{Technische Universit\"{a}t Wien}, \orgaddress{\street{Stadionallee 2}, \city{1020 Vienna}, \country{Austria}}}

\abstract{
Entanglement\textsuperscript{\cite{EinsteinPodolskyRosen1935}}, a key resource of emerging quantum technologies, describes correlations between particles that defy classical physics.
It has been studied extensively on various platforms\textsuperscript{\cite{FreedmanClauser1972, Giustina-Zeilinger2015, Hensen-etal-loophole-free2015}}, but has remained elusive in electron microscopy. 
Transmission electron microscopes are well-established tools for materials characterisation\textsuperscript{\cite{ReimerKohl2008, knoll_elektronenmikroskop_1932}} with unparalleled spatial resolution\textsuperscript{\cite{ishikawa2023spatial}}. They provide control over the preparation and detection of high energy electrons, with largely unexploited potential in the study of many-body quantum correlations.
Here, we demonstrate entanglement in electron-photon pairs generated via cathodoluminescence in a transmission electron microscope.
Employing coincidence imaging techniques adapted from photonic quantum optics\textsuperscript{\cite{bennink2004ghost_EPR, DAngeloKimKulikShih2004}}, we reconstruct both near- and far-field ``ghost'' images of periodic transmission masks. 
By measuring spatial and momentum correlations, we show a violation of the classical uncertainty bound: $\Delta x_-^2 \Delta k_+^2 = 0.502 \pm 0.047<1$.
Hence, we demonstrate entanglement in position and momentum\textsuperscript{\cite{rembold2025El_Ph_pair_entanglement}} -- the continuous variables at the base of most imaging methods, bridging the fields of electron microscopy and quantum optics. 
Our work paves the way for exploring quantum correlations in free-electron systems and their application to quantum-enhanced imaging techniques on the nanoscale.
}
\vspace{-5cm}
\maketitle

A state is said too be entangled when multiple particles are so strongly linked that they cannot be described as individuals\textsuperscript{\cite{NielsenChuang2000}}. 
This leads to correlations in non-commuting properties of the particles---such as position and momentum---that go beyond classical physics, regardless of the distance between them.
Since its initial conceptualization, entanglement has been extensively studied both theoretically and experimentally, with several key experiments conducted on photon pairs\textsuperscript{\cite{bellEinsteinPodolskyRosen1964, FreedmanClauser1972, Giustina-Zeilinger2015, Hensen-etal-loophole-free2015}}.

Advancing in parallel with the first quantum revolution, electron-based imaging methods have made significant technological progress since the first electron microscopes\textsuperscript{\cite{knoll_elektronenmikroskop_1932}}.
These advancements are reflected by the capabilities of transmission electron microscopy~(TEM)\textsuperscript{\cite{ReimerKohl2008}}, a highly developed technology that employs electron optics and the wave properties of electrons to resolve structures on the atomic scale\textsuperscript{\cite{ishikawa2023spatial}}, as well as reaching temporal resolution in the femto- and even atto-second range\textsuperscript{\cite{barwick_photon-induced_2009, baum2007attosecond, priebe_attosecond_2017}}. Consequently, TEM has become an indispensable tool in many fields of research, including materials science and biology.
Although established TEM techniques such as diffraction and holography\textsuperscript{\cite{tanaka_convergent-beam_2011, yoon_decoding_2024, lichte_electron_2007}} rely on interference phenomena, advanced quantum effects based on entanglement remain largely unexplored in this context.
Recently, entanglement in the TEM has gained the field's attention\textsuperscript{\cite{henke_ropers2025entanglement, Konecna2022entanglement_electron-photon, Kfir2019entanglement_electron-electron, kazakevich_kfir2024entanglement, rembold2025El_Ph_pair_entanglement,Schattschneider2018Entanglement, adiv_observation_2023, abajo_roadmap_2025}}, but so far direct experimental verification of its presence is lacking. 

Extrapolating quantum imaging techniques into electron microscopy will improve sensitivity beyond the shot noise limit, thus reducing radiation damage\textsuperscript{\cite{slussarenko2017unconditional_entangled}}, improve resolution\textsuperscript{\cite{rosi_increasing_2024}}, and potentially allow imaging with undetected probe particles\textsuperscript{\cite{Zou1991, Lemos2014ZWM_undetected_photons}}. 
Massive relativistic electrons have de Broglie wavelengths on the order of a few picometers, making them ideally suited for probing matter at the atomic scale. Photons, on the other hand, have been central to the development of quantum technologies\textsuperscript{\cite{bellEinsteinPodolskyRosen1964, FreedmanClauser1972, Giustina-Zeilinger2015, 2009photonic_QT_Vuckovic}} and come with a mature ecosystem of tools and techniques to manipulate and detect them. 
Hybrid electron–photon systems thus hold great promise for imaging, but also for generating entangled free electrons mediated by photonic channels\textsuperscript{\cite{henke_ropers2025entanglement}}.

Ghost imaging is an established technique from photonic quantum optics\textsuperscript{\cite{PittmanShihStrekalovSergienko1995, bennink2004ghost_EPR, DAngeloKimKulikShih2004, moreau2019ghost_imaging}}.
It uses the correlation between two photons to image an object on one optical path, while the detector is placed on the other. 
Classical systems can only be \textit{perfectly} correlated in one of two conjugate observables like position and momentum at a time, limited by the Heisenberg uncertainty principle.
This restriction on the correlations is not present in entangled states, thus allowing for ghost images to achieve diffraction-limited resolution in two joint conjugate measurement bases at once. 
Ultimately, these limitations of classical ghost imaging provide criteria for verifying quantum entanglement\textsuperscript{\cite{mancini_entangling_2002, DAngeloKimKulikShih2004}}.

In a TEM, photons are produced via cathodoluminescence~(CL), which is categorised as either coherent or incoherent\textsuperscript{\cite{2010deAbajo}}. 
Coherent CL, generated via the electrodynamics of a charged particle interacting with a dielectric medium, is of particular interest in entanglement studies because of the strong correlations between electron-photon pairs\textsuperscript{\cite{rembold2025El_Ph_pair_entanglement}}. 
The short timescales involved in coherent CL emission allow for the temporal filtering of incoherent CL contributions, which typically exhibit longer lifetimes\textsuperscript{\cite{Scheucher2022}}. With the advent of time-resolved single electron detectors\textsuperscript{\cite{poikelaTimepix365KChannel2014}}, coincidence-matching of single electrons with single optical photons has become possible, as demonstrated in\textsuperscript{\cite{ahn1985excited, varkentina_excitation_2023, yanagimotoTimecorrelatedElectronPhoton2023, feistCavitymediatedElectronphotonPairs2022}}. 
Such advances have inspired theoretical proposals on exploring quantum correlations and entanglement in transmission electron microscopy\textsuperscript{\cite{Konecna2022entanglement_electron-photon, henke_ropers2025entanglement, kazakevich_kfir2024entanglement, ben_hayun_shaping_2021, kfir_optical_2021, Kfir2019entanglement_electron-electron, rembold2025El_Ph_pair_entanglement}}.
Here, we utilise coincidence matching between electron-photon pairs to perform ghost imaging in both position and momentum space, to demonstrate that the joint measurement exceeds the classical limits set by Heisenberg's uncertainty principle for separable states.

\section*{Entanglement}
Entanglement is a type of correlation that goes beyond what is possible in classical physics\textsuperscript{\cite{HorodeckiEntanglementReview2009}}. 
Transition radiation and other coherent CL processes\textsuperscript{\cite{stoger2017transition, preimesberger_exploring_2025}} naturally lead to correlation in position and anti-correlation in momentum. Considering no other interactions, the photon is produced at the position $x_\gamma$ where the electron passes through the material, leading to $x_-=x_\text{e}-x_\gamma=0$ for each pair, where $x_\text{e}$ is the position of the electron. Similarly, coherent CL results in anti-correlation in momentum as the electron is deflected by the amount that the photon is created with $k_+=k_\text{e}+k_\gamma=0$, where $k_{\gamma}$ and $k_\text{e}$ denote the $x$-components of the photon and electron wavevectors, proportional to their momenta. 
Considering the fundamental limitations of wave mechanics, however, an electron with a localised position must be broadly distributed in momentum and vice versa, resulting in the Heisenberg uncertainty principle $\Delta x_\text{e}^2 \Delta k_\text{e}^2\geq 1/4$\textsuperscript{\cite{GriffithsQM2005}}. 
The same is true for photons.
Hence, the only way an electron and a photon can be perfectly correlated in position and anti-correlated in momentum is if their wave functions are not separable. 
That is, if the full wave function cannot be formulated as a statistical sum of products of electron and photon wave functions.
Here lies the core of the argument: non-separability is what defines entanglement\textsuperscript{\cite{Werner1989}}. 
Using the formal definition for separable states (see Ref.~\cite{BertlmannFriis2023}) and the Heisenberg uncertainty principle, one can show that a state must be entangled if it follows the relationship
\begin{equation}\label{eq:MGVT}
    \Delta x_-^2 \Delta k_+^2<1,
\end{equation}
also known as the Mancini-Giovannetti-Vitali-Tombesi inequality\textsuperscript{\cite{mancini_entangling_2002, duan_inseparability_2000}} (see Supplementary Information (SI): \hyperref[app:theory]{Entanglement Bound} for more details).
Thus, if the correlations in position and momentum are sharper than allowed for the two particles separately, the two particles must be entangled.
This formulation is state-agnostic, i.e., independent of the shape of the underlying distributions, as it depends only on their variances.  
Furthermore, it shows the entanglement in the continuous variables of position and momentum, rather than discretising the state space. 

\section*{Experimental section}

\begin{figure*}[htb]
	\centering
	\includegraphics[width=\textwidth]{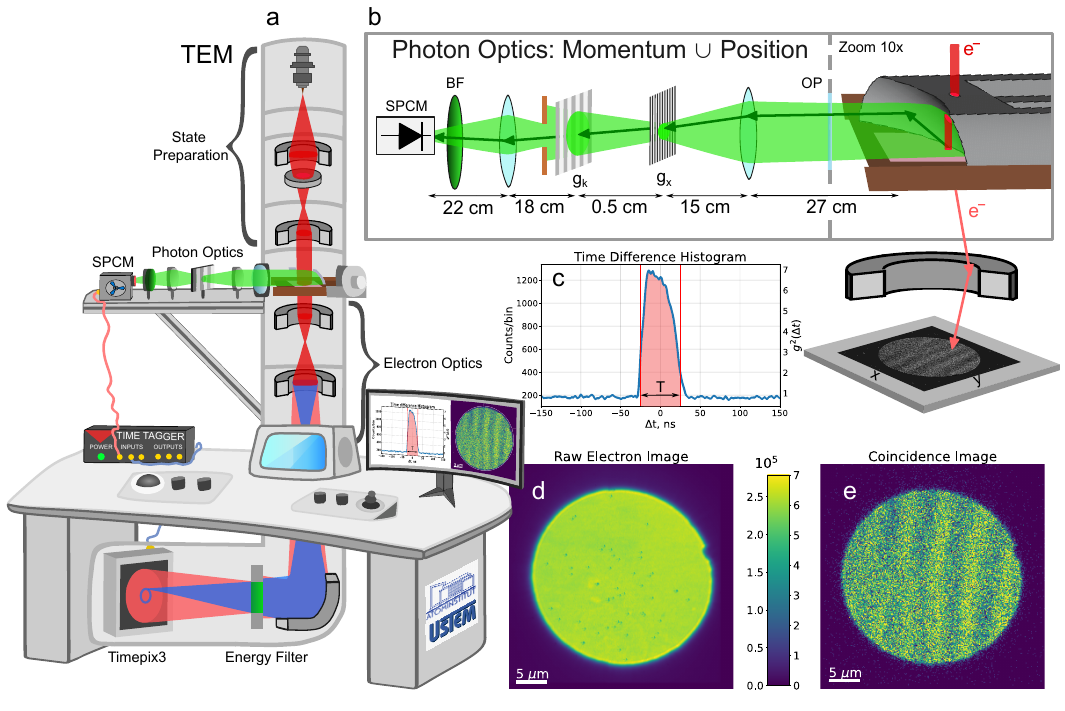} 
	\caption{\textbf{Experimental setup for detecting correlated electron-photon pairs in momentum and position space.} a) Schematic view of the adapted TEM. A 200~keV electron beam interacts with a thin monocrystalline Si sample, generating correlated electron-CL photon pairs. The transmitted electrons are energy-filtered around the loss corresponding to the emitted CL photon energy and detected, either in diffraction or position mode, using a Timepix3 camera, which timestamps each event. Simultaneously, the corresponding CL photons are collected by a parabolic mirror which is mounted on the sample holder and directed through an optical viewport (OP) to the photon optics. b) The optical setup allows for the insertion of an absorptive grating mask either in momentum ($g_k$~=~100 \si{\micro\meter}) or image space ($g_x$~=~38.5 \si{\micro\meter}). Chromatic and spherical aberration contributions are reduced using a bandpass filter centred at 550~nm (BF) and an aperture. The transmitted photons are detected and timestamped by a single photon counting module (SPCM). A time tagger ensures the necessary synchronization, enabling the temporal correlation of individual photons with their originating electrons. c) This correlation is visible in a $g^{(2)}$ plot. d) Raw electron beam image recorded by the Timepix3 camera. e) Coincidence image reconstructed from correlated electron-photon pairs, revealing the structure of the grating mask. 
        }
	\label{fig:setup}
\end{figure*}

Fig.~\ref{fig:setup} gives a general overview of the experimental setup we used to probe momentum and position correlations in electron–photon pairs, allowing direct access to the joint uncertainties.
We use an adapted TEM (FEI Tecnai G2 F20) with a primary electron beam energy of $E_\mathrm{kin}= 200\; \mathrm{keV} \pm 0.45~\mathrm{eV\;(HWHM)}$ to irradiate a $50~\mathrm{nm}$ thick monocrystalline silicon membrane with a highly collimated electron beam (diameter $\sim 23 ~\mu$m). 

Upon interaction with the sample, each electron may produce a CL photon. These photons originate almost exclusively from transition radiation and are known to be anti-correlated in momentum with their emitting electron\textsuperscript{\cite{preimesberger_exploring_2025}}, thus allowing for ghost imaging. 
A custom-built parabolic mirror mounted on the TEM sample holder collects the photons emitted from the sample. The paraboloid is aligned to collect photons from the beam position and deflect them through an optical view port on the side of the TEM column to our photon optics system shown in Fig.~\ref{fig:setup}b. 
Inserting a lens in the photon path forms a spatially resolved image of the electron beam -- the source of the CL radiation -- on the sample, i.e. photons emitted from one point on the sample are mapped onto one point in the image plane. 
For ghost imaging, we insert a resolution test target, consisting of periodic opaque lines on a transparent substrate into the image plane, thus either transmitting or blocking photons depending on their position. 

In order to investigate the momentum of the emitted CL photons, we place the mask in the Fourier plane of the imaging system, thus mapping photons onto the mask according to their momentum. 
Subsequently, we use an additional lens to collect the transmitted photons and guide them to a single-photon counting module (SPCM) or image the CL photons onto a low noise CMOS camera for alignment of the photon optics.
Before reaching the SPCM, a bandpass filter selects photons with a wavelength range of $550\;\mathrm{nm} \pm 20 \;\mathrm{nm}$ (half-width at half maximum: HWHM). Each photon detection event is timestamped by a time tagging device.

Within the TEM, the electrons transmitted through the sample are further guided by a series of adjustable magnetic lenses, allowing for two detection options: 
in imaging mode, the magnetic lenses are set to form an image of the transmitted wave function at the detector plane, resulting in a position measurement of the electron (Fig.~\ref{fig:setup}d); while in diffraction mode, the back focal plane of the objective lens is projected onto the detector, thus implementing a momentum measurement of the electron. 
Prior to detection, electrons pass through a Gatan GIF 2001 energy filter, which selects those that have undergone an energy loss matching the CL photon emission. 
Finally, electrons are recorded by a TimePix3 direct electron detection camera, which timestamps each event. Both electron- and photon-detector are synchronised using the time tagging unit. 
By analysing the arrival times of the collected detection events, we identify coincident electron-photon pairs. As illustrated in Fig.~\ref{fig:setup}e), this step reveals the coincidence image of the object placed in the photon's path. 

We bound the joint uncertainties $\Delta x_-^2$ and $\Delta k_+^2$ by comparing the coincidence data $I(x,y)$ in position space ($J(k_x,k_y)$ in momentum space) to the expected image in the detector plane (see Methods:~\hyperref[ssec:Analysis]{Data Analysis}). 
Initially, the mask is modelled as a binary image $\mathcal{M}(x',y')$ in position ($\mathcal{M}'(x',y')$ in momentum).
The effective mask on the specimen plane $\mathcal{M}_\mathrm{eff}(x,y)$ (or $\mathcal{M}'_\mathrm{eff}(k_x,k_y)$) is distorted by the parabolic mirror and can be retrieved by taking the setup's geometric properties into account. 

The measured coincidence image is modelled by a convolution of this effective mask with a Gaussian point spread function (PSF), $\mathcal{G}(\sigma_x, \sigma_y)$ (and $\mathcal{G}(\sigma_{k_x}, \sigma_{k_y})$).
We fit the PSF to the data, minimising the least absolute error $\mathrm{LAE}_x$ (and $\mathrm{LAE}_k$), to extract the standard deviations $\sigma_x,~\sigma_y$ (and $\sigma_{k_x}, ~\sigma_{k_y}$):

\begin{equation}
\begin{split}
\mathrm{LAE}_x&=|I(x,y)  - I_0(x,y)\cdot[\mathcal{M}_\mathrm{eff}*\mathcal{G}(\sigma_x, \sigma_y)](x,y)|,\\
\mathrm{LAE}_k&=|J(k_x,k_y)  - J_0(k_x,k_y)\cdot[\mathcal{M}_\mathrm{eff}*\mathcal{G}(\sigma_{k_x}, \sigma_{k_y})](k_x,k_y)|,
\end{split}
\end{equation}

where $I_0$ (and $J_0$) is the intensity distribution in position (momentum) of coincidence electron-photon pairs measured without a mask in the photon path. The resulting standard deviations of the PSF estimate the average inference error over the entire detection window $\Delta_\text{infer} x,\;\Delta _\text{infer}k$, further discussed in the Methods:~\hyperref[ssec:Analysis]{Data Analysis}. The inference error describes the certainty of one particle's measurement outcome when provided with a measurement of the other\textsuperscript{\cite{ReidDrummondBowen2009}}. It provides an upper bound for the desired position and momentum uncertainties: 

\begin{equation*}
   \Delta x_-\leq\Delta_\text{infer} x,\quad \Delta k_+\leq\Delta_\text{infer} k.
\end{equation*}

\section*{Results}

\begin{figure}
	\centering
	\includegraphics[width=90mm]{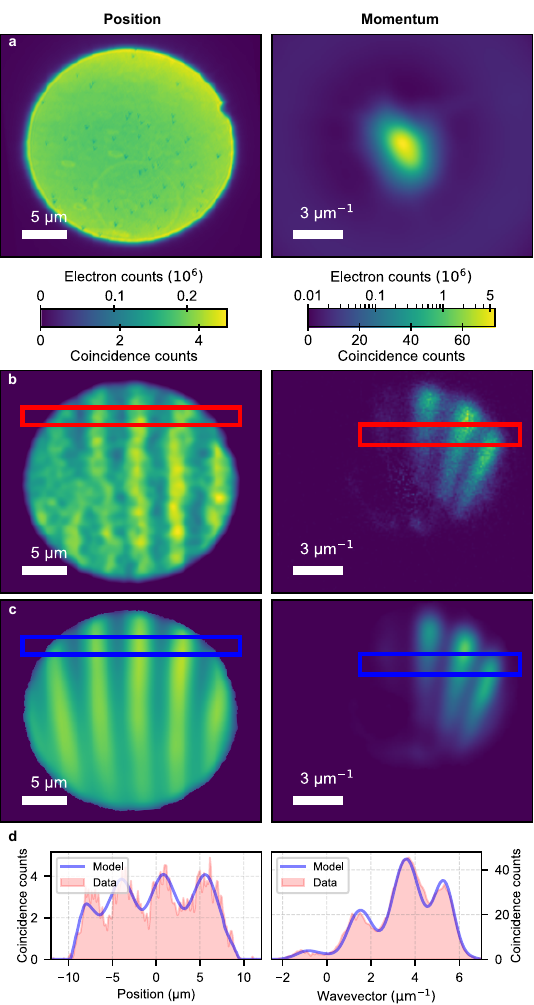}
	\caption{\textbf{Experimental data in position (left column) and momentum (right column) space, used for entanglement verification.} a) shows the unprocessed electron distributions. Temporally correlating these electrons with photons transmitted through a grating mask reveals the coincidence images depicted in b). The models in c) are obtained by convolving the expected shape of the photon mask, after accounting for mirror distortions, with a Gaussian PSF. d) shows line profiles through the regions marked by the red and blue squares in b) and c). We applied a smoothing filter to the position data in b) (raw data shown in Extended Data Fig.~\ref{fig:residuals}a) and rotated the images for illustrative purposes. The direction of the line profile in position space thus doesn't coincide with the x-direction in the analysis.
    }
	\label{fig:data}
\end{figure}

We obtained position and momentum datasets, each containing $N>10^5$ coincidence events, Fig.~\ref{fig:data}. All experiments were performed in an uninterrupted 30 hour session. Calibration datasets without a mask in the photon beam path were recorded for both settings, thus measuring $I_0,~J_0$ (also $>10^5$ coincidence counts). 

Fig.~\ref{fig:data} shows the recorded data in position space in the left column, and in momentum space on the right. Temporal correlations of electrons and photons reveal coincidence images of the inserted photon masks, shown in Fig.~\ref{fig:data}b. The parabolic mirror distorts the image of the grating lines projected back onto the specimen plane. We calculate these distortions (details in SI:~\hyperref[ssec:geometry]{Geometric Distortion Mapping}) and account for the unmodulated distributions $I_0,~J_0$ shown in Extended Data Fig.~\ref{fig:data_mixed}a,b. In momentum space, the features in the $J_0$ dataset reflect the limits imposed by the parabolic mirror and the aperture used in the photon path. In position space, the features of the $I_0$ distribution are determined by the condenser aperture in the electron probe-forming system.
Using these in our model, we achieve excellent agreement with the experimental images, therefore the resulting model depicted in Fig.~\ref{fig:data}c which captures the defining features of the recorded data. A discussion on the residuals can be found in SI:~\hyperref[ssec:residuals]{Model Residuals}.

The position data exhibits a joint uncertainty of $\Delta x_- \leq 1.448 \pm 0.061~$µm, while the corresponding momentum measurement yields $\Delta k_+ \leq 0.489 \pm 0.010 ~$µm$^{-1}$. This results in an overall uncertainty product of
\[\Delta x^2_-\Delta k^2_+  \leq 0.502 \pm 0.047<1,\]
which falls below the bound for separable states and provides a verification of entanglement between the electron and its corresponding CL photon by $\sim10$ standard deviations.

The error on this value is determined by the limited number of coincidence counts as well as systematic errors, such as the limited accuracy of position and momentum calibration in the TEM, and the hysteresis in the TEM's magnetic lenses, as detailed in the Methods section and the SI. 

In order to validate the correct choice of photon position $x_\gamma$ and momentum $k_\gamma$ measurement bases, we additionally recorded smaller mixed-basis $k_\gamma,x_\mathrm{e}$ and $x_\gamma,k_\mathrm{e}$ datasets ($\sim 3\cdot10^4$ coincidences), shown in Extended Data Fig.~\ref{fig:data_mixed}c,d. As expected, the recorded data in conjugate measurement bases exhibits no ghost images, confirming independence of position and momentum bases.

\section*{Conclusion}
Using ghost imaging techniques adapted from photonic quantum optics, we confirm, with high statistical confidence, entanglement between single electrons and their associated CL photons. We prove this via measurements of the position and momentum correlations of the electron-photon pair, demonstrating that the product of their variances lies below the threshold obtainable with separable states. Hence, these pairs cannot be described classically. 
Although our experiments were conducted using transition radiation, our conclusions can be readily extended to other types of coherent CL, such as Cherenkov radiation\textsuperscript{\cite{rembold2025El_Ph_pair_entanglement}}. 

Apart from the Gaussian PSF used to estimate the variance, our methods are independent of the underlying wavefunctions. 
Our experimental approach can be extended to facilitate detailed reconstruction of the quantum state, by e.g., measuring the full correlation images to directly extract the average inference error\textsuperscript{\cite{ReidDrummondBowen2009, achatz_certifying_2022}}.
Furthermore, measurements in mutually unbiased bases\textsuperscript{\cite{rembold2025El_Ph_pair_entanglement}} could be used to discretise the continuous variables, thus enabling the use of advanced quantum information tools developed for discrete-variable systems\textsuperscript{\cite{NielsenChuang2000}}.
This work marks a crucial step toward integrating well-established photonic quantum optics techniques\textsuperscript{\cite{Zou1991, Lemos2014ZWM_undetected_photons, 2009photonic_QT_Vuckovic, moreau2019ghost_imaging}} with the powerful platform of electron microscopy\textsuperscript{\cite{ishikawa2023spatial, abajo_roadmap_2025}}.

\bibliography{electron-photon-entanglement}
\bibliographystyle{sn-nature}

\section*{Methods}
\subsection*{Experimental Setup}
\subsubsection*{Photon Optics}\label{sssec:photonoptics}
To measure the spatial and momentum distribution of CL photons we built a free-space optical setup. A custom-designed parabolic mirror, with a focal length of approximately \SI{750}{\micro\meter}, is mounted directly onto the TEM specimen holder. The monocrystalline Si membrane is positioned near the focal point of the mirror, allowing the mirror to efficiently collect the emitted CL photons and collimate them into a nearly parallel beam directed out of the TEM and into the optical system.

To produce high-quality CL photon images, we use an achromatic doublet lens (anti-reflection coated for 400–700 nm) with a focal length of 10 cm. This configuration results in a 9.5$\times$ magnified image of the CL beam formed 147 mm downstream of the lens, defining the first image plane. To perform coincidence position measurements, an absorptive grating mask (Thorlabs R1L3S6P) is used as the object for ghost image formation.
This target features lithographically-produced chromium oxide line gratings of varying periods, fabricated on a transparent soda lime glass substrate. 
The opaque chrome pattern is specified to have an optical density larger than three. 
The transmissivity of the transparent portion is specified to be above 90\%, limited by reflections at the air-glass and glass-air interfaces. 
It is mounted on a translation stage to ensure precise and reproducible positioning at the first image plane. For coincidence momentum measurements, the grating mask is positioned in the Fourier plane, located 5 mm from the first image plane, using the same translation stage. To perform position and momentum correlation measurements two different grating periods are used: a finer grating ($g_x$ = 38.5  \si{\micro\meter}) for position space coincidence imaging and a larger grating period ($g_k$ = 100 \si{\micro\meter}) for momentum measurements. 

To further guide photons transmitted through the mask, a second identical lens is placed 180 mm from the mask, positioned in the first image plane. To reduce chromatic aberrations, we employ a bandpass filter ($550\;\mathrm{nm} \pm 20 \;\mathrm{nm\;(HWHM)}$). A flippable planar mirror placed after the lens allows the beam to be redirected toward either a single-photon counting module (PicoQuant PMA Hybrid 40 mod) for correlation measurements or a CMOS sensor for alignment of the optical system and the grating mask.
The CMOS sensor is positioned 219 mm from the second lens, capturing an image with a magnification of 1.2:1 relative to the actual physical size of the mask to approximately maintain the original size of the CL beam.

For the alignment of the optical system we use the scanning TEM (STEM) mode, with high electron beam current. We focus the converged electron beam onto the Si membrane to produce highly localised CL emission. The resulting emitted photons are collimated by the parabolic mirror, pass through the optical system and are detected, ideally as diffraction limited ``points", by the CMOS sensor. Alignment of the photon optics is optimised by adjusting the optical components to minimise the spot size and maximise the brightness of the CL signal on the CMOS image. 

To align the specimen stage with the optical axis of the optical system, we use a scanning electron beam to expose the sample at predefined locations, simultaneously acquiring images of the CL spots with the CMOS camera (CL maps). The acquisition time of the CMOS camera is synchronised with the dwell time of the electron beam. This procedure results in a spatially resolved discrete CL map, which reveals the quality of our imaging optics as a function of emission location on the sample. By analysing this map, we identify the region corresponding to the mirror surface with the highest spatial resolution. 

Subsequently, the microscope is switched to TEM mode and the electron beam is once again focused onto the sample surface. The specimen stage is then adjusted such that the position of the CL spot observed in STEM mode aligns precisely with the CL spot recorded in TEM mode.
This registration process - acquiring a CL map in STEM mode and repositioning the specimen stage in TEM mode — is iteratively repeated until the CL spot remains fixed on the camera sensor, indicating proper alignment in TEM mode.

Once the specimen stage is aligned, an aperture positioned in the parabolic mirror's image plane selects the region with optimal optical performance, reducing contributions from spherical aberrations and imperfections of the parabolic mirror. To achieve this, the CMOS camera is translated relative to the last lens to image the parabolic mirror surface. The aperture is then translated until its image becomes visible on the camera, indicating that it is positioned in the image plane of the parabolic mirror surface. After this alignment, we shift the CMOS back to the image plane of the CL beam, where the CL spot size is minimised and the brightness of the CL signal is maximised on the CMOS image. The aperture is gradually closed until the CL spot exhibits minimal aberrations. This limits the active area of the parabolic mirror used for image formation, effectively reducing its numerical aperture, while improving the overall optical performance of the system.

To align the grating mask in the position plane, we expand the electron beam to a diameter of approximately 40 \si{\micro\meter} and insert the grating mask into the optical path of the photons. For coarse alignment, the mask position is adjusted until a sharply focused image appears on the CMOS camera. For momentum coincidence measurements, the grating mask is positioned in the Fourier plane, which, according to ray diagram construction, is located approximately 5 mm from the position image plane. Fine alignment of the grating mask in position or momentum space is performed during electron-photon coincidence measurements. In this step, the grating is shifted in $\sim$100 \si{\micro\meter} increments to maximise the contrast of the resulting coincidence image.

\subsubsection*{Electron Optics}
Our TEM is operated at an accelerating voltage of 200 keV, with a beam current of approximately 10 pA.
To optimise our incident electron beam parameters, we adjust the TEM's minicondenser lens, a small lens immediately before the upper objective lens, in coordination with the second condenser lens and the upper objective lens, to produce a parallel beam on the sample with a diameter of approximately 23~\si{\micro\meter} and an electron momentum resolution $\sim500$~\si{\nano\radian} (estimated by the FWHM of the primary beam after the Si membrane, at the relevant energy loss).
The TEM is operated in low magnification mode, which sets the objective lens to a very low current value. This configuration allows us to achieve the extremely long diffraction camera length, required to resolve the small-angle (on the order of a few \si{\micro\radian}) scattering from coherent CL photon emission.

Following the projection system, the electron beam enters a Gatan GIF 2001 energy filter, where we filter the electrons by energy to isolate those which have lost energy to the production of an optical photon. We use a 1~\si{\electronvolt} wide energy slit centred at an energy loss of $2.8~\text{eV}$. The image or momentum plane of the filtered electron beam is projected onto a Timepix3-based event-counting detector (Advascope ePix) mounted at the TV port of our spectrometer. These energy slit parameters provide a compromise between reducing the background from the high number of counts from unscattered electrons, and providing a good coincidence matching rate with the 550~\si{\nano\meter} ($\widehat{=}~2.25~\text{eV}$) filtered photons.

In position space, the system operates at a magnification of $487\times$, and in momentum space, we operate at a camera length of 1.5 \si{\kilo\meter}.

\subsubsection*{Acquiring Coincidence Images}
The electron detector outputs a series of data packages containing the position (i.e. the pixel index), time of arrival and time over the threshold of each detection event\textsuperscript{\cite{poikelaTimepix365KChannel2014}}. The clocks of the electron and photon detectors are synchronised by a time tagging module (Swabian Instruments Time Tagger Ultra), which outputs timestamps for each detected photon and ensures synchronisation with the electron detector. Our electron-photon coincidence setup has a timing resolution of approximately 50 \si{\nano\second} \textsuperscript{\cite{preimesberger_exploring_2025}}.
 
The coincidence data acquisition pipeline used for these experiments is largely unchanged from that described in\textsuperscript{\cite{preimesberger_exploring_2025}}, with the exception of electron event clustering and the drift correction routine. In contrast to our previous work, the clustering step has been omitted in this investigation. As our coincidence matching algorithm is set up to identify at most one coincidence count for each photon, we find no significant deterioration in coincidence imaging performance under the low current measurement conditions presented in this work.

Over the course of the experiment, the electron beam may shift due to environmental disturbances or instabilities. Our post-processing routine accounts for these drifts via a drift correction. Contrary to our previous publication, we now apply a separate correction for each data package containing approximately $650\,$k electron detection events. The difference between the beam position and the centre of the detector (pixel coordinate 127, 127) is calculated and a corresponding shift is applied to all events in the package. For the position measurement we take the beam centre to be the centre of mass (COM) of the electron detection events. 

For momentum, however, the COM calculation is disproportionately influenced by scattering of electrons into the light cone\textsuperscript{\cite{preimesberger_exploring_2025}}, near the edges of the camera. At the large camera lengths we are using, the electron beam in momentum space is particularly sensitive to disturbances. Even small deflections can shift part of the light cone scattering off the camera and distort the COM calculation. Instead, we aggregate the electron events into an image and apply a Gaussian blur ($\sigma = 2$ pixels) followed by a simple local maximum peak-finding method offered by the scikit-image Python library\textsuperscript{\cite{waltScikitimageImageProcessing2014}}).

Two other minor changes from the data processing procedure described in\textsuperscript{\cite{preimesberger_exploring_2025}} are a change in the centre of the interval chosen for determining false coincidence counts from \SI{-100}{\nano\second} to \SI{-150}{\nano\second}, and a change of the expected time delay $\mathbb{E}[\Delta t_\mathrm{e\gamma}]$ to \SI{-40}{\nano\second} due to new cabling. 

We obtained approximately $10^5$ coincidence counts for both position and momentum measurements at a rate of $\sim$7 counts per second, requiring over four hours of acquisition per dataset. Reference images without grating lines were also acquired yielding similar count numbers but at twice the rate, consistent with the grating’s transmission properties. 
Additionally, two mixed-basis datasets, photon position - electron momentum and photon momentum - electron position, were recorded ($\sim 30,000$ counts each), to validate the correct implementation of the measurement bases, see Extended Data Fig.~\ref{fig:data_mixed}.

The microscope is located on the eighth floor in downtown Vienna, in close proximity to metro and tram lines. To minimize external disturbances, data acquisition was carried out during late-night and weekend hours. Disturbances to the measurement typically manifest as energy shifts in the electron spectrometer and we continually monitor the count rate during acquisition and adjust the energy shift as needed.

\subsection*{Data Analysis}\label{ssec:Analysis}

\subsubsection*{Fitting}

In order to fit the shape of the mask as projected onto the sample, we create a model describing the transformations induced by our optical system. The mask shape $\mathcal{M}(x',y')$ ($\mathcal{M}'(x',y')$ for momentum) is modelled as binary and is fully described by its period $g_i$ with $i\in
\{x,k\}$ and offset $\ell_i$. 

\begin{align*}
    \mathcal{M}(x',y',g_i,\ell_i) &= \begin{cases} 
      0 & (x'+\ell_i)\mod{g_i} < g_i/2\\
      1 & (x'+\ell_i)\mod{g_i} \geq g_i/2\\
    \end{cases}
   \label{eq:pos_mask},
\end{align*}
with $x',y'$ referring to the position on the surface of the mask.
Considering a binary mask, although a continuous profile is expected in the setup, ensures that all blurring is accounted for in the inference error, preventing its underestimation.

To obtain $\mathcal{M}_\mathrm{eff}(x,y)$ (and $\mathcal{M}'_\mathrm{eff}(k_x,k_y)$), we calculate the mapping $\mathcal{M}(x',y')\to \mathcal{M}_\mathrm{eff}(x,y)$ (and $\mathcal{M}'(x',y')\to \mathcal{M}'_\mathrm{eff}(k_x,k_y) $) corresponding to the setup's geometric optics (see SI:~\hyperref[ssec:geometry]{Geometric Distortion Mapping} for more details) and the conversion between real space and position (and wavevector) in the image plane. 

The Gaussian convolution in Eq.~\eqref{eq:fit} is inspired by equivalent techniques from photonic quantum optics\textsuperscript{\cite{courme_manipulation_2023}} based on spontaneous parametric down-conversion. However, ideally, the PSF should be adapted to the shape of the correlation distribution. In our case, the convoluted model reproduces the shape of the data very well. The coefficient of determination is $R^2=0.47$ in position and $R^2=0.92$ in momentum space. Finally, the data does not exhibit tails or additional maxima which would indicate a PSF with a larger variance.

In the mapping we assume that the transformation is purely geometrical, resulting in effective masks that are still binary. Under experimental conditions, aberrations, small misalignments, or contamination on the optics may introduce further setup-inherent blurring and unknown deformations. If the model perfectly represented the image, the fitting procedure would not need to compensate for small shape mismatches. In this ideal case, it would minimise the fitted variance and find the true uncertainties. As our model is not perfect, we expect that the variances are effectively overestimated and bound the correlation uncertainties from above. For a discussion of the residuals, see SI:~\hyperref[ssec:residuals]{Model Residuals}.

In this setup, coincidence imaging shows the intensity variations experienced by the photons. The transmission mask that is imaged effectively modulates the distribution $I_0$ ($J_0$) that would be present without the mask. Since $I_0$ ($J_0$) is not homogeneous (Extended Data Fig.~\ref{fig:data_mixed}a,c), we recorded a flat-field correction data set of $\sim10^5$ coincidence counts. 
The accuracy of the flat-field correction is mainly limited by shot noise. Therefore, we apply a smoothing algorithm to obtain the unmodulated intensity distributions.

First, we apply a median filter to the raw image to remove the effects of stray electrons outside the primary beam profile. Since the primary beam is smaller than the field of view of the electron camera, zero counts or hot pixels are often registered around the edges. To define the shape of the beam, we apply a threshold to the median-filtered image and create a binary mask.
After applying this mask to the raw data, we iteratively apply Gaussian filters with $\sigma=5\,$pixels for position and $\sigma=2\,$pixels for momentum to non-zero pixels. Including zero-valued pixels would artificially lower the intensity around the edges. We conclude the iterative smoothing with a Gaussian filter that includes contributions from the camera edges to avoid an unphysically abrupt cut-off.

The geometrical mappings are fully described by 10 parameters: the periodicities $g_x$, $g_k$;  the focal length of the parabolic mirror $f$; the angles of the coordinate system rotations due to the electron optics $\phi_x$, $\phi_k$; the distance $d_\mathrm{Focus}$ between the sample holder and the first lens outside the microscope; the 3D position of the electron beam with respect to the focal point $\vec{x}_\mathrm{Beam}$; and finally the offsets $\ell_x$, $\ell_k$ of the gratings.

$g_x, g_k$ and $f$ are precisely known and fixed in the fitting process.  

The relative position of the illuminated area with respect to the focal point $\vec{x}_\mathrm{Beam}$, as well as $\ell_x$, $\ell_k$ are not known. While our alignment procedure guarantees the optimal imaging conditions of the beam on the sample plane, we cannot measure its coordinates directly. Hence, we fit $\vec{x}_\mathrm{Beam}$ along with the standard deviations of the PSF. The same is true for the grating offsets $\ell_i$ as they are inextricably linked to the beam position. 

$\phi_x,~\phi_k$ and $d_\mathrm{Focus}$ are approximately known from separate calibration measurements. During analysis, we observe a persistent rotation of the measured intensity distribution by $\sim$5 degrees with respect to the measured values for both $\phi_x$ and $\phi_k$. We consider this to be a systematic offset and fixed the expected rotation value in the model to account for it during fitting. For $d_\mathrm{Focus}$, we found that the value of best fit quality deviates from the measured by about 3\%, we considered this to be within the margin of error for our manual measurement and fixed this parameter to the corrected value. 

As we include the global parameters $\vec{x}_\mathrm{Beam}, \ell_i$ in the fitting function, the least absolute errors for position and momentum need to be jointly evaluated to yield consistent results and avoid over-fitting, we thus minimise their sum: 

\begin{equation}
    \underset{\sigma_x, \sigma_y, \sigma_{k_x},\sigma_{k_x}, \vec{x}_\mathrm{Beam},\ell_i}{\arg\min} (\mathrm{LAE}_x + \mathrm{LAE}_k)
\label{eq:fit}
\end{equation}

The LAEs are evaluated for each pixel and summed for the full dataset. The optimisation is carried out using the Nelder-Mead algorithm\textsuperscript{\cite{nelderSimplexMethodFunction1965}} included in the SciPy library\textsuperscript{\cite{2020SciPy-NMeth}}. 

To ensure the correct integral, i.e. number of counts, for the model, it is normalised with respect to the target data.  

The choice of fixed parameters does not significantly impact the fitting outcome, when $\phi_x,~\phi_k, ~d_\mathrm{Focus}, ~\vec{x}_\mathrm{Beam}$ and $\ell_i$ are included as variables in the fitting routine, the result for $\sigma_x,~\sigma_{k_x}$ and their errors remain practically unchanged, see the SI for more detail.

\subsubsection*{Error Estimate}

In order to gauge an upper bound on the statistical error on our estimates of $\Delta_\text{infer}x$ and $\Delta_\text{infer}k$, we randomly subsample our full dataset, which contains $\sim10^5$ valid coincidence counts for each measurement. We draw 20 subsamples without replacement, each containing 25,000 coincidence counts, and evaluate the fit described in Eq.~\ref{eq:fit} for each of them. 

The noise in the unmodulated intensities $I_0, J_0$ is negligible after smoothing and corresponds to the best representations of the unmodulated beam shapes. Hence, they are not resampled for each batch.  

The variance in the Gaussian widths over all subsampled produces the following upper bounds for the statistical errors:  

\begin{equation*}
    \Delta\sigma_{x\;\mathrm{Stat}}<0.059\;\mathrm{\mu m}, \quad \Delta\sigma_{k_x\;\mathrm{Stat.}}<0.006\;\mathrm{\mu m}^{-1}
\end{equation*}

Systematic errors are caused by the limited accuracy of position and momentum calibration in the TEM, as well as hysteresis. As we need to switch between the position and momentum basis over the course of the experiment, the magnetic fields produced by the lenses may deviate slightly from the calibration measurement. These errors summarize to $\Delta\sigma_{x\;\mathrm{Cal}}\approx 0.017\;\mathrm{\mu  m}$ and $\Delta\sigma_{k_x\;\mathrm{Cal.}}\approx 0.007 \;\mathrm{\mu  m^{-1}}$. See the SI:~\hyperref[ssec:calibration]{TEM Calibration} for more details on the calibration procedure and the impact of hysteresis on our calibrations. 

\section*{Data Availability}
The datasets generated and analysed during the current study are available from the corresponding author upon reasonable request.

\section*{Code Availability}
The code used for analysis during the current study are available from the corresponding author upon reasonable request.

\section*{Funding Information and Acknowledgments}
The authors thank Michael Scheucher, Mario Krenn, Dominik Hornof, Santiago Beltrán-Romero, Elizabeth Agudelo, Nicolai Friis, Dennis R\"{a}tzel and the USTEM team for support and fruitful discussions. PR thanks Martin Bohmann for methodological discussions. PH, AP, SB, IB thank the Austrian Science Fund (FWF): Y1121, P36041, P35953. This project was supported by the ESQ-Discovery Program 2020 "A source for correlated electron-photon pairs" and the FFG-project AQUTEM. PR acknowledges funding from the Austrian Federal Ministry of Education, Science and Research via the Austrian Research Promotion Agency (FFG) through the flagship project FO999897481 (HPQC).
\\

During preparation of this manuscript, we became aware of an independent research project on related aspects of electron-photon entanglement, in the group led by Claus Ropers.

\section*{Author Contributions}

P.H. conceived the initial experiment and directed the study. A.P., S.B., I.B. and P.H. designed the experiment, made the measurements and carried out the data analysis with the help of P.R.. P.R. and A.P. adapted and developed the theoretical framework with the help of the other authors. All authors contributed to the manuscript.

\section*{Competing Interests}
The authors declare no competing interests.

\section*{Additional Information}
Supplementary information is available for this paper.

Correspondence and requests for materials should be addressed to Philipp Haslinger at philipp.haslinger@tuwien.ac.at.

\newpage 
\numberwithin{figure}{section}
\setcounter{figure}{0}
\renewcommand{\thefigure}{\arabic{figure}}
\renewcommand{\figurename}{Extended Data Fig.}
\section*{Extended Data}
\subsection{Conjugate bases measurements}
\begin{figure*}[htb]
	\centering
	\includegraphics[width=\textwidth]{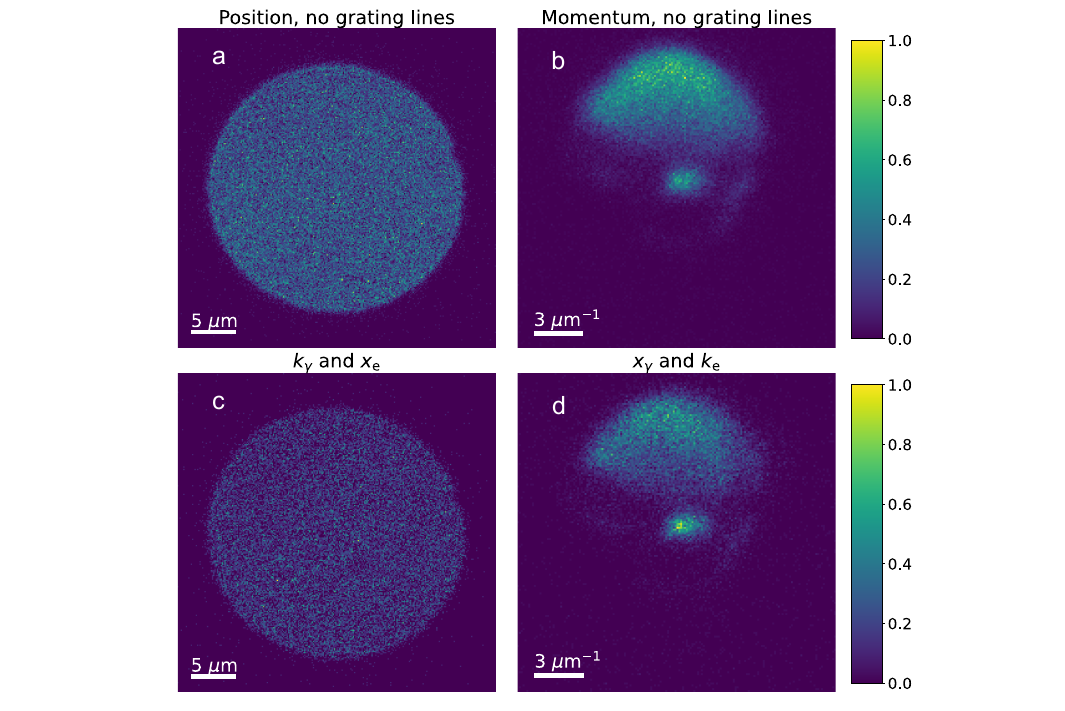} 
	\caption{\textbf{Recorded experimental data in electron position and momentum bases with no grating lines and with grating lines in conjugate bases.} a) Reference coincidence image with no grating lines in the electron position basis (with the TEM operated in image mode). b) Reference coincidence image with no grating lines in the electron momentum basis (with the TEM operated in diffraction mode). A ghost image of a small region of the parabolic mirror can be observed. c) Coincidence image in the mixed basis: photon momentum $k_\gamma$ (with the grating mask placed in the Fourier plane) and electron position $x_\mathrm{e}$. d) Coincidence image in the other mixed basis: photon position $x_\gamma$ (with the grating mask placed in the first image plane) and electron momentum $k_\mathrm{e}$. Each image was normalised by maximum intensity.
    }
	\label{fig:data_mixed}
\end{figure*}

\clearpage
\newpage
\subsection*{Model Residuals}

\begin{figure*}[htb]
    \centering
    \includegraphics[width=70mm]{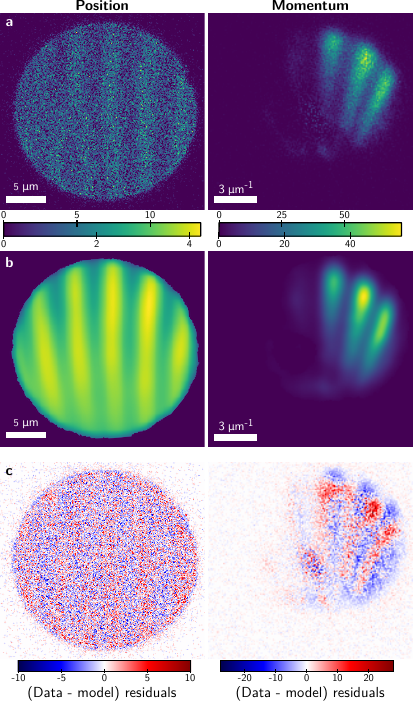}
    \caption{\textbf{Residuals from the fitted model.} Left column contains data for position space, right column contains data for momentum space. a) Raw coincidence data, as presented in Fig.~\ref{fig:data}, with no smoothing applied. b) Fitted models, repeated from Fig.~\ref{fig:data} for convenient visualisation. c) Fit residuals, obtained by subtracting the model from the data. Colourbars are in units of coincidence counts.}
    \label{fig:residuals}
\end{figure*}

\setcounter{equation}{0}
\renewcommand{\theequation}{S.\arabic{equation}}

\clearpage
\newpage
\section*{Supplementary Information}

\subsection*{Entanglement Bound}~\label{app:theory}

The Mancini-Giovannetti-Vitali-Tombesi inequality\textsuperscript{\cite{mancini_entangling_2002}} used to witness entanglement in this work reads
\begin{equation}
    \Delta x_-^2 \Delta k_+^2 < 1,
\end{equation}
where variances are defined as $\Delta f^2=\langle (\Delta f)^2\rangle$.
In the following we give an illustrative derivation to explain its origins.
Let us start by noting that for any single system $j\in\{A,B\}$, we can formulate the Heisenberg uncertainty principle:
\begin{equation}
    \Delta x_j^2 \Delta k_j^2 \geq \frac{1}{4}.
    \label{eq:HUP}
\end{equation}
To understand how it relates to witnessing entanglement in entangled bipartite states, we first define their opposite: separable states:
\begin{equation}
    \rho_\text{sep}=\sum_s P_s \;\rho_{A,s} \otimes \rho_{B,s}.
\end{equation}
They are representable as a convex sum of tensor products of pure states $\rho_{j,s}$ with the distribution $P_s$. 
Equivalently, the joint probability distributions for a variable $f$ can be expressed as
\begin{equation}
    P(f_A,f_B)=\sum_s P_s \;P(f_A|s) \cdot P(f_B|s).
\end{equation}

Before coming to the derivation, let us note the following useful relationships: 
First, for positive real numbers
\begin{equation}
     \frac{a}{b}+\frac{b}{a} \geq 2 \;\leftrightarrow\; a^2+b^2\geq 2\sqrt{a^2b^2}.
    \label{eq:posreal}
\end{equation}
Secondly, the Cauchy-Schwartz inequality
\begin{equation}
    \left(\sum_s P_s \Delta f_s^2\right)\left(\sum_s P_s\Delta g_s^2\right)\geq\left| \sum_s P_s \sqrt{\Delta f_s^2 \Delta g_s^2}\right|^2.
    \label{eq:CauchySchwarz}
\end{equation}
Finally, the variance of any sum of distributions of a variable $f$ follows the relationship\textsuperscript{\cite{mallon_bright_2008}}
\begin{align}
    \Delta f^2 \geq \sum_s P_s \Delta f_s^2. 
    \label{eq:variance}
\end{align}

Consequently, for separable states, we assume that $\Delta x_{-,s}^2=\Delta x_{A,s}^2+\Delta x_{B,s}^2\geq 2\sqrt{\smash[b]{\Delta x_{A,s}^2\Delta x_{B,s}^2}}$ and $\Delta k_{+,s}^2=\Delta k_{A,s}^2+\Delta k_{B,s}^2\geq 2\sqrt{\smash[b]{\Delta k_{A,s}^2\Delta k_{B,s}^2}}$ using Eq.~\eqref{eq:posreal}. 
Multiplying the two variances and applying the relationships above we get the following condition for separable states:
\begin{align*}
    \Delta x_-^2 \Delta k_+^2 &\geq \sum_s P_s \Delta x_{-,s}^2\sum_s P_s \Delta k_{+,s}^2\\
    &\geq \left| \sum_s P_s \sqrt{\Delta x_{-,s}^2 \Delta k_{+,s}^2}\right|^2\\
    &\geq \left| \sum_s P_s \;\sqrt{4\;\sqrt{\Delta x_{A,s}^2\Delta x_{B,s}^2 \;\Delta k_{A,s}^2\Delta k_{B,s}^2}}\right|^2\\
    &\geq \left| \sqrt{4\sqrt{\frac{1}{4^2}}}\right|^2 = 1.
\end{align*}
Thus we recover the bound given at the start of this section.
A more thorough version of this derivation is provided in the original work\textsuperscript{\cite{mancini_entangling_2002}} starting at Eq.~(10).

\subsection*{Deriving Model Parameters}\label{ssec:modelparameters}
\subsubsection*{Optical Magnification}
Using the acquired CL maps, we determine the optical magnification of the CL beam. Since the step size of the scanning electron beam is predefined, the spatial period of the CL maps can be measured and used to calculate the effective optical magnification of the system, which includes contributions from the parabolic mirror and the two lenses.

However, because the second lens is located downstream of the grating lines and does not contribute to the formation of the ghost image, its magnification must be excluded from this calculation. During the alignment procedure described in the \hyperref[sssec:photonoptics]{photon optics} section, two final CL maps were acquired with predefined electron beam step sizes of 3.6 \si{\micro\meter} and 6 \si{\micro\meter}. The measured spatial periods of the corresponding CL maps, excluding the optical magnification from the second lens, were 34.4 \si{\micro\meter} and 56.6 \si{\micro\meter}, respectively.

From these values, the optical magnification of the CL beam was determined to be $9.50 \pm 0.11$.

\subsection*{Model Residuals}\label{ssec:residuals}

To test the goodness of our fit, we present in Extended Data Fig.~\ref{fig:residuals} the raw data, fit, and residuals. The residuals are obtained by subtracting the model from the data. 

Further, we calculate the coefficient of determination ($R^2$ value) of our fit using the formula

\begin{equation}
    R^2 = 1 - \frac{\sum_i(x_i - m_i)^2}{\sum_i (x_i - \bar{x})^2},
\end{equation}

where $x_i$ represents the intensity at each pixel in the raw data, $m_i$ the intensity at the corresponding pixel in the model, and $\bar{x}$ the mean value of the raw data. For momentum space, we obtain $R^2 = 0.92$, and for position space, $R^2=0.47$. 

The lower $R^2$ value of the position fit is due to the higher contributions from Poissonian shot noise to the signal, which causes high variance from pixel to pixel. The model fits the whole image and cannot account for the high frequency fluctuations caused by the relatively low number of coincidence counts per pixel. Although both images were collected with a similar total number of coincidence counts, in the momentum dataset, the counts are concentrated on a much smaller number of pixels on the camera, and so the signal-to-noise ratio is much higher. For the position data, the counts are spread over many pixels and so contributions from shot noise are clearly visible in the raw data. The smooth model cannot account for the high variance from the Poissonian noise. 

Nevertheless, examining the trend in the residuals, we observe that the position fit is underestimating the peaks in the raw data and overestimating the signal in the valleys, indicating that the $\sigma_x$ value we have obtained from the fit is an underestimation of the true distribution.
We also observe on the left side of the raw data that there is a blurred patch in the position image. This blurred patch is likely due to an aberration or smudge in the photon optics. Its contribution to the raw image is fully included in the fitting routine, but not in the model, and will contribute to increasing the net $\sigma_x$.

The momentum data has a higher signal to noise ratio, with more counts per pixel, but there are still some effects from the presence of shot noise in the residuals. The structure in the residuals comes from imperfections in the correspondence between the model and the `true' projection of the mask onto the sample's momentum plane.

\subsection*{Geometric Distortion Mapping}\label{ssec:geometry}
Both the position and momentum mapping are obtained using ray optics and by modelling our mirror as an ideal, rotationally symmetric paraboloid with a focal length \mbox{f~=~\SI{750}{\micro\meter}}. 
For the momentum model, we compute the point on the mirror's Fourier plane corresponding to a given wavevector $\vec{k}$. We do this by finding the intersection point of a ray travelling from the sample position $\vec{x}_\mathrm{Beam}$ in direction $\vec{k}$ and being reflected by the mirror, with a ray travelling in the same direction, starting from the focal point. The projection from the Fourier plane onto the grating mask $\mathcal{M}'$ is modelled as an ideal image transformation with a lens positioned at $d_\mathrm{Focus} =317\;\mathrm{mm}$ from the mirror's focal point (f = \SI{100}{\milli\meter}). This corresponds to a distance between the first lens and the image plane of $\sim$\SI{143.7}{\milli\meter}, deviating from the manually measured value of \SI{147}{\milli\meter} by \SI{3.3}{\milli\meter} (2.3\%). We consider the fitted value to be more reliable and thus use it as a fixed parameter for the model.  
Similarly, the position projection model is obtained by geometrically determining the virtual image of the photon position in the ideal parabolic mirror. This is also done by computing the virtual intersection point of a ray passing through $\vec{x}$, travelling parallel to the optical axis with a ray passing through both $\vec{x}$ and the mirror's focal point. We then model the projection onto the grating mask $\mathcal{M}$ as an ideal image transformation using the same lens parameters.
Modelling the influence of the imaging system in this way allows us to capture the large scale image distortions due to the reflection from the parabolic surface while retaining a binary mask.

In order to validate, our choice of fixed vs. fitted parameters, we evaluated the result for the alternative analysis approach of including all the parameters which are not known precisely into the fitting routine, thus evaluating
\begin{equation}
    \underset{\sigma_x, \sigma_y, \sigma_{k_x},\sigma_{k_x}, \vec{x}_\mathrm{Beam},\ell_i,\phi_x,\phi_k, d_\mathrm{Focus}}{\arg\min} (\mathrm{LAE}_x + \mathrm{LAE}_k),
\end{equation}, 
following the same subsampling procedure described in the error estimation section. The resulting values are qualitatively equivalent:

\begin{align*}
\Delta k \mathrm{\;[\mu m^{-1}]}&\leq 0.489 \pm 0.011\\
\Delta x \mathrm{\;[\mu m]}& \leq  1.448 \pm 0.031\\
\Delta x^2\cdot \Delta k^2 &\leq 0.502 \pm 0.031
\end{align*}

The following table summarized the obtained parameter values:  

\begin{center}
\begin{tabular}{ c| c c c}
& Mean & Std. Dev. & relative Error\\
\hline
$\phi_x$ & 30 \si{\degree} &  0.29\si{\degree}& 1.0\%\\
$\phi_k$ & 295 \si{\degree} & 0.22\si{\degree}& 0.1\%\\
$x_\mathrm{Beam}$& -2.39 \si{\micro\meter} & 0.05 \si{\micro\meter} & 2.2\%\\
$y_\mathrm{Beam}$& -32.6 \si{\micro\meter} & 0.12 \si{\micro\meter} & 0.4\%\\
$z_\mathrm{Beam}$&  52.4 \si{\micro\meter} & 0.14 \si{\micro\meter} & 0.3\%\\
$\ell_x$ & 34.6 \si{\micro\meter} & 0.29 \si{\micro\meter} & 0.8\%\\
$\ell_k$ & $3.8 \cdot 10^{-4}$ \si{\micro\meter} & $8.4 \cdot 10^{-6}$ \si{\micro\meter}  & 2.2\%\\
$d_\mathrm{Focus}$ &317.0 mm & 0.3 mm & 0.1\%\\
\end{tabular}
\end{center}

\subsection*{TEM Calibration}\label{ssec:calibration}
To verify the accuracy of our electron optics magnifications under custom alignment settings, we calibrated the TEM magnification using a standard diffraction replica sample from Agar Scientific. This sample features shadow-cast carbon replicas of diffraction line gratings with a known spacing of 462.9 nm and is commonly used for calibration of TEM image and diffraction space. Immediately after completing the acquisition of the coincidence data, we removed our TEM holder and inserted the calibration sample in a single tilt holder. Calibration data were then acquired using the same beam settings as were used for data acquisition, ensuring consistency between the experimental and calibration conditions.

\subsubsection*{Momentum}
For the momentum calibration, we acquired ten images of the diffraction pattern from the 462.9 nm grating, using the same settings as in the electron momentum measurements. Between each calibration data set, the electron beam was slightly shifted (on the camera) to sample different regions of the imaging system. In this way, angular-dependent aberrations are incorporated into the final magnification calibration and reflected in the error bars. In four of the images, two diffraction spots were visible, while the remaining six images displayed four spots. We fit a Gaussian to each spot and extract the $x$ and $y$ location of the centre of the Gaussian with sub-pixel precision. From these fits, we calculated the distance in pixels between neighbouring spots and retrieved 28 measurements used for momentum calibration.

The square grid produces diffraction spots at $2\pi/\lambda_{sq}=13.57$~\si{\micro\meter^{-1}}. From our 28 measurements, we derive our calibration of $8.95 \cdot 10^{-2}$~\si{\micro\meter^{-1}}/pixel with a standard deviation of $\sigma= 0.13\cdot10^{-2}$~\si{\micro\meter^{-1}}/pixel across our measurements.

\subsubsection*{Position Calibration}
Using the same diffraction grating replica, we took four images at the same TEM magnification as was used for the electron position measurements, integrating $1000$ frames with an exposure time of 1 \si{\milli\second} per frame.

We take the Fourier transform of each of these images, each yielding 13 reflections in Fourier space. For each reflection, we fitted a two-dimensional Gaussian and extracted the peak positions with sub-pixel precision. As in the momentum calibration, we computed the distances between neighbouring peaks in pixels.
Over all four images, this resulted in 64 measurements between Fourier points. Using the known grating spacing of 462.9 \si{\nano\meter}, we calibrate the Fourier-transformed images and invert this calibration to calibrate the real-space images. The final result shows a mean value of 112.9 \si{\nano\meter}/pixel, with a standard deviation of $\sigma=1.3$ \si{\nano\meter}/pixel.

\subsubsection*{Hysteresis}
Due to the magnetic nature of the TEM's lenses, they suffer from hysteresis when switching between modes. Hysteresis in the lenses can impose some variation in the magnification on the electron camera. To explore these effects, we switched between position and momentum and back ten times, taking images of the same calibration sample with each iteration, using the same settings as for our experiment. These values give an indication of how much the magnification calibration can change during our experiment as we switch back and forth between position and momentum space.

For position space, we have 19 images, with 2 images for all but one iteration. We perform the same calibration procedure as we used for the magnification calibration. For each image, we identify 5 spots in the FFT, yielding 4 measurements. In total, we have 10 data points for nine of the iterations, and 5 points for the iteration with a single image. We then take the average per loop and calculate the standard deviation across these ten measurements to identify how the hysteresis could impact our calibration. This procedure results in a standard deviation of $0.26\cdot10^{-3}$ \si{\nano\meter}/pixel.

Likewise, in momentum space, we have 20 images, two for each iteration of our hysteresis test. Of these images, all but one have four diffraction spots from the 462.9 nm grating in; the remaining image has two spots. From these images, we obtain 8 distance measurements for each iteration (5 for the iteration with the image with two spots). We calculate the standard deviation of the means for each iteration: $1.97\cdot 10^{-4}$ \si{\micro\meter^{-1}}/pixel.
This procedure was performed several days after the initial experiment.

\end{document}